\documentstyle[12pt]{article}
\input{epsf}
\begin{document} 
\begin{center}
\Large\bf{$^{56}$Ni dredge-up in Supernova~1987A}
\end{center}
\vspace*{0.5cm}
\large{A.Fassia$^{1}$ and  W.P.S. Meikle$^1$\\
\\ 
\small
$^1$Astrophysics Group, Blackett Laboratory, Imperial College, Prince Consort Rd
, London SW7 2BZ, UK\\
\begin{abstract}
We use early-time observations of He~I 10830~\AA\ to measure the extent
of upward mixing of radioactive material in SN~1987A.  This work
develops and extends the work of Graham (1988), and places constraints
on actual explosion models.  The presence of the He~I 10830~\AA\
(2s$^{3}$S--2p$^{3}$P) line at $\geq10$~days post-explosion implies
re-ionisation by $\gamma$-rays from upwardly-mixed radioactive material
produced during the explosion.  Using the unmixed explosion model~10H
(Woosley 1988) as well as mixed versions of it, we estimated the
$\gamma$-ray energy deposition by applying a purely absorptive
radiative transfer calculation. The deposition energy was used to find
the ionisation balance as a function of radius, and hence the 2s$^{3}$S
population density profile. This was then applied to a spectral
synthesis model and the synthetic spectra were compared with the
observations.  Neither model~10H nor the mixed version, 10HMM,
succeeded in reproducing the observed He~I 10830~\AA\ line.  The
discrepancy with the data found for 10HMM is particularly significant,
as this model has successfully reproduced the X-ray and $\gamma$-ray
observations and the UVOIR light curve.  We find that a match to the
He~I line profile is achieved by {\it reducing} the extent of mixing in
10HMM.  Our reduced-mixing models also reproduce the observed
$\gamma$-ray line light curves and the iron-group velocities deduced
from late-time infrared line profiles. We suggest that the He~I line
method provides a more sensitive measure of the extent of mixing in a
type~II supernova explosion.

\end{abstract}
\hspace*{0.3cm}
\vspace*{0.2cm}
\newpage

\section{Introduction}
It has long been recognised (Falk \& Arnett 1973) that
hydrodynamic instabilities should occur in type~II supernova
explosions. Chevalier (1976) studied the stability of shocks
propagating through simple power-law density distributions and showed
that for some cases the matter behind the shock is subject to
instabilities. Bandiera (1984) further emphasised the role of such
instabilities in generating chemical mixing.

SN~1987A provided the first observational evidence for extensive mixing
of the ejecta.  Simple stratified models did not account for the
extended, smooth, rounded plateau in the UVOIR light curve which was
observed during the era covering 25 to 125 days post-explosion.  To
successfully model the light curve, it proved necessary to
invoke upward mixing of radioactive nickel and cobalt to high
velocities together with downward mixing of hydrogen deep into the
core (Arnett 1988, Woosley 1988, Shigeyama {\it et al.} 1988,
Shigeyama \& Nomoto 1990, Nomoto {\it et al.} 1998).  Upward mixing of
radioactive $^{56}$Ni and $^{56}$Co to the outer regions of the
envelope was also indicated by the early detection of X-rays ({\it
e.g.} Dotani {\it et al.} 1987, Sunyaev {\it et al.} 1987) and
$\gamma$-rays ({\it e.g.} Matz {\it et al.} 1988, Mahoney {\it et al.}
1988, Sandie {\it et al.} 1988, Cook {\it et al.} 1988). Furthermore,
the expansion velocities inferred from the line widths of infrared
spectral lines of Fe~II and Co~II (Spyromilio, Meikle \& Allen 1990,
Haas {\it et al.} 1990, Tueller {\it et al.} 1990) indicated that a
fraction of the iron-group elements had been dredged up to velocities
as high as $\sim$3000~km/s.

These observations stimulated the modelling of explosions both in two
dimensions (Arnett {\it et al.} 1989, Hachisu {\it et al.} 1990,
Fryxell {\it et al.} 1991) and three dimensions (M\"{u}ller {\it et
al.} 1991). This theoretical work showed that significant mixing occurs
as the explosion shock plows through the ejecta. However, the maximum
iron-group velocities derived did not exceed 1300~km/s. This
discrepancy with the observations was attributed to non-inclusion in
the models of the radioactive decay energy deposited in the
ejecta. Herant \& Benz (1991) included the radioactive decay energy in
their two-dimensional code, resulting in the formation of a giant
``nickel bubble'' during the first few weeks.  In spite of this, their
model still only attained iron-group velocities up to $\sim$2000~km/s -
1000~km/s less than that indicated by the observed iron-group line
profiles. 
 
Herant \& Benz (1992) excluded the possibilities that the velocity
discrepancy was due to three-dimensional effects, errors in the adopted
progenitor structure, or numerical resolution. They argued that in
order to explain the observed high velocities, it was necessary to
invoke mixing during the early stages of the explosion when nickel is
created by explosive nucleosynthesis (Herant {\it et al.} 1994).  By
{\it pre-mixing} nickel out to 1.5 M$_{\odot}$ above the mass cut, they
found that the subsequent Rayleigh-Taylor instabilities could produce
iron-group velocities as high as those observed.  Such pre-mixing is
possible within the context of the delayed explosion mechanism (Herant,
Benz, \& Colgate (1992); Herant {\it et al.} 1994).  In this scenario,
the explosion is driven by neutrinos from the proto-neutron star which
deposit energy in material just above the neutrinosphere, producing a
hot, high-entropy bubble interior to low-entropy shocked matter.  This
situation is prone to strong convective motions under the influence of
the gravitational pull of the proto-neutron star, thus breaking the
spherical symmetry. The residual inhomogeneities due to these early
instabilities may be sufficient to seed subsequent shell-interface
Rayleigh-Taylor instabilities.

Another way to account for the observed high velocities is through the
effect of a spherically symmetric shock on irregularities or ``seeds''
formed in the  $^{16}$O burning shell before core-collapse. Recent
hydrodynamic calculations (Bazan \& Arnett 1998, Arnett 1994) have
shown that the oxygen burning is intermittent, chaotic and strongly
localised. The heterogeneous composition developed provides ``seed''
perturbations in density. These are believed to be sufficiently large
to produce hydrodynamic instabilities which cause mixing in the
presupernova envelope. This occurs in precisely the region where
$^{56}$Ni is explosively produced by oxygen burning behind the
explosion shock. However, quantitative results are not yet available.

To test the possible mixing mechanisms described above we need to be
able to probe the conditions deep inside the supernova. In principle,
much can be learned from the $\gamma$-ray and X-ray light curves.  In
addition, once the ejecta have become optically thin (t$\geq150$ days),
high signal-to-noise infrared spectral line profiles of heavy elements
can provide a powerful probe of their spatial distribution, and degree
of mixing (Spyromilio, Meikle \& Allen 1990, Herant \& Woosley 1994).
Unfortunately, apart from the exceptionally close SN~1987A, current
instruments are insufficiently sensitive to allow such studies to be
made for typical nearby type~II supernovae.

An interesting alternative is to probe the mixing of the nickel using
infrared spectra at early times. This approach was first presented by
Graham (1988), and was developed further by us (Fassia {\it et al.}
1998). The method is based on the detection and modelling of the
He~I~10830~\AA\ (2s$^{3}$S--2p$^{3}$P) line. This is a high excitation
line.  It is formed in a region where the population of the metastable
2s$^{3}$S level is maintained by the balance of the recombination rate
of He$^{+}$ with the rates of collisional de-excitation and forbidden
radiative decay. However, because the He$^{+}$ recombination timescale
is $<$~1~d (Graham 1988) we expect that at late times (t~$>$~10~d) all
helium should be neutral and in its ground state. Thus, if
He~I~10830~\AA\ is present at times after 10~d there must exist a
re-ionising mechanism.  The identification of this line well after 10~d
in SNe 1987A and 1995V led Graham (1988) and Fassia {\it et al.} (1998)
to propose that the ionisation was maintained by the $\gamma$-rays
emitted from the decay of $^{56}$Ni/$^{56}$Co. The subsequent
recombination produces the He~I 10830~\AA\ line. Owing to the
exponential sensitivity of the $\gamma$-ray flux to the column depth
(Pinto \& Woosley 1988) the observed He~I line profile is sensitive to
the amount of radioactive material moving at high velocities.  Thus,
the He~I 10830~\AA\ line can probe the extent of upward mixing or
``dredge-up'' of iron-group elements in a type~II supernova.  This was
demonstrated by Fassia {\it et al.}  (1998) for the type~IIp supernova
SN~1995V. While an unmixed explosion model did not reproduce the
observed He~I line, by invoking upward mixing of about 10$^{-6}$
M$_{\odot}$ of $^{56}$Ni to velocities above 4000 km/s (i.e. above the
helium photosphere) we were able to match the observed spectra.

In the present work we use the He~I line method to carry out a detailed
study of the dredge-up of radioactive material in the ejecta of SN
1987A, and to place constraints on possible mixing models.  The method
is described in section~2.  In section~3 we present results from the
comparison of theoretical models with the data and estimate the
$^{56}$Co dredge-up. We also compare our predictions of $\gamma$-ray
fluxes with the observed $\gamma$-ray light curves.  In section 4 we
discuss the implications of this work in constraining and understanding
the dredge-up.

\section{Determination of $^{56}$Ni dredge-up in SN~1987A using the 
He~I 10830~\AA\ line}
In spectra of SN~1987A obtained in the period 76-135 days
post-explosion, Elias {\it et al.} (1988) and Meikle {\it et al.}
(1989) identified a strong absorption feature at $\sim$10,700~\AA\ with
the blue-shifted P~Cygni trough of He~I 10830~\AA. Using the spectral
synthesis model described in Fassia {\it et al.} (1998), where the
lines are formed by pure scattering above the photosphere and where
line blending is taken into account, we have confirmed this
identification and shown that the He~I line was blended with
C~I~10695~\AA. The He~I line trough was blueshifted by about 5000~km/s.
 This is in sharp contrast to other infrared P~Cygni lines where
velocities of less than 2000~km/s were observed (Elias {\it et al.} 1988,
Meikle {\it et al.} 1989). The high blueshift and depth of the trough
indicate that the line was optically thick, and formed in the outermost
layers of the expanding supernova envelope.

As explained above, the presence of the He~I line implies re-ionisation
by the $\gamma$-rays of $^{56}$Co.  Modelling of the He~I line profile
and evolution was carried out as follows. We began with the unmixed
explosion model 10H (Woosley 1988), calculated the radioactive
energy deposited in the outer envelope of the supernova and hence found
the ionisation balance as a function of radius. We then determined the
2s$^{3}$S population density profile.  This was fed into our spectral
synthesis model and the synthetic line profile was compared with the
observations. As will be shown below, model 10H drastically
underproduced the He~I 10830~\AA\ feature, suggesting that upward
mixing of radioactive material must have occurred during the explosion.
To determine the degree of this dredge-up we gradually increased the
mixing of the $^{56}$Co until a match with the observations was
obtained.

\subsection{Calculation of the energy deposition and the emergent
luminosity of the $\gamma$-rays }

$\gamma$-ray energy deposition was computed by performing a radiation
transport solution for each $\gamma$-ray line produced in the decay
$^{56}$Co$\rightarrow$$^{56}$Fe. Line energies and emission
probabilities were obtained from Lederer \& Shirley (1978). We also
included $\gamma$-rays from the annihilation of positrons emitted in
19\% of the $^{56}$Co decays. For this reason, following Leising \&
Share (1990), we used a branching ratio
of 0.38 for $\gamma$-rays with energies of 0.511~MeV.

We estimated the $\gamma$-ray energy deposition and the escape
fractions, using a purely absorptive transfer equation.  While Monte
Carlo simulations provide the most realistic and physically accurate
description of the $\gamma$-ray energy deposition, they are excessively
demanding of computational resources.  In contrast, the radiative
transfer method, adopted in this study, is computationally efficient
and has been shown to give results reasonably close to those obtained
by the Monte Carlo techniques (Swartz {\it et al.} 1995).  This is
because the $\gamma$-ray energies are such that Compton scattering
dominates.  The ratio of scattered to incident photon energies is given
by

\begin{equation}
\frac{E_{\nu}^{\prime}}{E_{\nu}}=\frac{1}{1+(E_{\nu}/m_{e}c^{2})(1-\cos\theta)}
\end{equation}
where $E_{\nu}$ is the initial photon energy, and $E_{\nu}^{\prime}$ is
photon energy after it scatters from an electron at rest through angle
$\theta$. 
Thus the energy transfered in a single scatter can be large {\it e.g.}
a $\sim$2 MeV $\gamma$-ray photon gives up $\geq$0.8 of its energy when
it scatters through $\theta\geq90^{\circ}$ (Swartz {\it et al.} 1995).
Even though the scattering cross section is forward peaked,
forward-scattering transfers very little energy to the gas.  Therefore,
one can think of the $\gamma$-ray line photon as travelling along a
linear path until it suffers a large-angle scatter, at which point it
gives up most of its energy and becomes a ``continuum''
photon. Consequently, in calculating the line transfer, the Compton
scattering can be treated as producing a purely absorptive opacity.
Thus, the radiative transfer equation is linear for the $\gamma$-ray
lines. We can therefore calculate the energy deposition and escape
intensities by direct integration of the radiative transfer
equation. The emissivity for this calculation is given by the local
volume rate of $\gamma$-ray production. Following Woosley's {\it et
al.} (1994) approach, the opacity is assumed to consist of scattering
and absorptive parts. The scattering component is

\begin{equation}
\kappa_{scat}^{\nu}=\frac{N_{e}\sigma_{KN}(E_{\nu})}{1+(E_{\nu}/m_{e}c^{2})
(1-\cos\theta_{scat})}
\end{equation}
and the absorptive part is:
\begin{equation}
\kappa_{abs}^{\nu}=N_{e}\sigma_{KN}(E_{\nu})[1-\frac{1}{1+(E_{\nu}/m_{e}c^{2})
(1-\cos\theta_{scat})}]
\end{equation}
where $E_{\nu}$ is the initial photon energy, $N_{e}$ is the total
(bound and free) electron density, $\sigma_{KN}(E_{\nu})$ is the
Klein-Nishina cross section, and $\theta_{scat}$ is the scattering
angle. In our calculations we have followed (Woosley {\it et al.} 1994)
in assuming that all photons scatter through the same angle,
$\theta_{scat}$=90$^{\circ}$. This assumption gives results that agree
to within a few percent with those obtained from more detailed
Monte-Carlo calculations.

The code solves the frequency dependent equations of radiative transfer
by integrating along impact parameters parallel to the observer's line
of sight (see Figure~1).  Following Swartz {\it et al.} (1995) the
transfer equation in the ``(p,z) representation'' along a ray is:
\begin{equation}
\pm \frac{\partial I^{\pm}_{\nu}}{\partial z}(p,z)=\eta_{\nu} -
\kappa_{\nu}(r) 
I^{\pm}_{\nu} 
\end{equation}
where $I^{+}$ ($I^{-}$) denotes the outgoing (ingoing) radiation, z is
the position along the ray (Figure~1) and
$\eta_{\nu}=f_{rad}s_{\gamma}(E_{\nu})\rho/4\pi$ is the local
$\gamma$-ray emissivity. Here, $f_{rad}$ is the initial mass fraction of
$^{56}$Ni, $\rho$ is the mass density and $s_{\gamma}(E_{\nu})$ is
the time-dependent rate of energy release per gram of radioactive
material:
\begin{equation}
s_{\gamma}(E_{\nu})=\frac{ 1.25\times
10^{17}E_{\nu} b_{E_{\nu}} e^{\frac{-t}{\tau_{Co}}}}{\tau_{Co}} (erg~g_{rad}^{-1}~s^{-1})  
\end{equation}
where $b_{E_{\nu}}$ is the branching ratio for a line of energy
$E_{\nu}$, and $\tau_{Co}$ is the mean life time of $^{56}$Co
($\tau_{Co}$=113.7 days).  Introducing the optical depth along the ray,
$d\tau=-\kappa_{\nu}dz$, the equation of transfer becomes:
\begin{equation}
\mp \frac{dI^{\pm}_{\nu}}{d\tau}=-\frac{\eta_{\nu}}{\kappa_{\nu}}+ I^{\pm}. 
\end{equation} 
If we now take into account that $N_{e}=(\rho/m_{u})Y_{e}$ where
$m_{u}$ is the atomic mass unit and $Y_{e}$ is the
total number of electrons per baryon, we have:
\begin{equation}
\mp \frac{d{\cal I}^{\pm}_{\nu}}{d\tau}=-\frac{f_{rad}}{Y_{e}}+ {\cal
I}^{\pm}  
\end{equation}
where 
\begin{equation}
{\cal I}=(\frac{4\pi\kappa_{\nu}}{s_{\gamma}m_{u}N_{e}})I.
\end{equation}
Since no radiation is incident from outside the ejecta the boundary
conditions are: $ I^{-}$=0 at the upper boundary and from symmetry at
$z=0$, the lower boundary condition is
$I^{-}(p,z)=I^{+}(p,z)$. Integration of equation (8) gives:
\begin{equation}
{\cal I}(\tau_{i})={\cal I}(\tau_{i+1})e^{-\Delta
\tau}+\int_{\tau_{i}}^{\tau_{i+1}} \frac{f_{rad}}{Y_{e}}e^{-(t-\tau_{i})}dt
\end{equation}
where $\tau_{i}$ and $\tau_{i+1}$ are the optical depths to points i,
i+1 along the ray and $\Delta\tau$=$(\tau_{i+1}-\tau_{i})$. The impact
parameter grid consists of rays tangent to concentric shells (Figure 1),
with a single ray passing through the center. Since $f_{rad}$, $Y_{e}$
are constant within each shell, equation (9) can be integrated exactly.

By setting
$\kappa_{\nu}=\kappa_{abs}^{\nu}$
we can calculate the rate ($R_{E}$) at which the energy is deposited
locally (in erg cm$^{-3}$ s$^{-1}$) from:
\begin{equation}
R_{E}=\sum_{\nu} 4\pi\kappa_{\nu} J_{\nu}
\end{equation}
where $J_{\nu}$ is the mean intensity of the $\gamma$-rays with initial
energy $E_{\nu}$.  The emergent luminosity of the individual
(unscattered) $\gamma$-ray line photons can also be calculated with the
$\gamma$-ray transport calculation described above by setting
$\kappa_{\nu}=N_{e}\times\sigma_{KN}(E_{\nu})$ (Woosley {\it et
al.} 1994).

\subsection{Calculation of the He~I 10830~\AA\ line}

In a H/He envelope of low ionisation, all the energy released by the
Compton scattering of the $\gamma$-rays is channelled into ionisation of
the H and He, rather than into heating the electron gas (Meyerott 1980).
Thus, we assume the key populating process for the helium levels is
recombination. We ignore direct excitation by fast electrons since they
only excite singlet states, which decay rapidly back to the ground
state. Of the recombinations to excited levels of He~I, approximately
three-quarters are to triplet states, with the remainder going to
singlet states (Osterbrock 1989).  Recombinations to singlet states
cascade rapidly to the ground state.  However, because there is a
substantial optical depth in the 1s$^{1}$S-np$^{1}$P transitions
(Graham 1988) all singlet recombinations will eventually pass through
the 2s$^{1}$S state. Atoms in the 2s$^{1}$S level decay by two-photon
emission (A=51~s$^{-1}$) to the ground state.  Therefore, only a small
population of excited singlet states is built up.

Recombinations to triplet states lead, through downward radiative
transitions, to the highly metastable 2s$^{3}$S level where the
population can become quite substantial. As described in Fassia {\it
et al.} (1998), there are a number of mechanisms which can depopulate
this level.  Firstly, a very weak single-photon
radiative decay to 1s$^{1}$S can occur ($A=1.27\times10^{-4}$
s$^{-1}$) (Osterbrock 1989).  Of considerably greater significance,
however, is the 2-stage intersystem radiative decay
2s$^{3}$S$\rightarrow$2p$^{3}$P$\rightarrow$1s$^{1}$S. This occurs in
the presence of a radiation field (such as from the photosphere),
which is required to excite the first stage. There are also two
important depopulating processes involving collisions. One of these
involves thermal electron collisions causing excitation or
de-excitation from 2s$^{3}$S across to singlet states.  The other
process is Penning ionisation (Bell 1970, Chugai 1991):
\begin {equation}
He~I~ (2^{3}S)+ H \rightarrow He~I~ (1^{1}S) + H^{+} + e.
\end {equation}
which takes place when hydrogen and helium are microscopically mixed. 
Therefore, the population balance of the 2s$^{3}$S 
level is described by the equation:
\begin{equation}
n_{He(2^{3}S)} (n_{e}Q + C_{P} + R + A)= \alpha(n^{3}L) n_{e}n_{He~II}.
\end {equation}
where $Q$ is the sum of the collision rates from 2s$^{3}$S to all singlet
states, and has the value 1.826$\times$10$^{-8}$cm$^{3}$s$^{-1}$ (Berrington
\& Kingston  1987). $C_{P}$ is the Penning ionisation rate and is given
by \begin{math} C_{P} =\gamma_{P} n_{H} \end{math} where
\begin{math}\gamma_{P}= 7.5\times 10^{-10}(\frac{T}{300K})^{1/2}
\end{math} cm$^{3}$s$^{-1}$ (Bell 1970). For the
recombination temperature of hydrogen (T$\sim$5000 K), which we adopted
for the days concerned (Catchpole {\it et al.} (1987)),
$\gamma_{P}$=3$\times$10$^{-9}$cm$^{3}$s$^{-1}$.   $\alpha(n^{3}L)$ =
3.26$\times$ 10$^{-13}$cm$^{3}$s$^{-1}$ is the total recombination
coefficient for the triplet states at a temperature of 5000~K
(Osterbrock 1989). A is the spontaneous transition probability and R is
the two stage inter-system radiative decay rate (Chugai 1991). This is
given by:
\begin{equation}
R(v)=B_{23}\frac{4\pi}{c}(\frac{D}{vt})^{2} F_{\nu}^{C}e^{0.92 A(\lambda)}
A_{32}^{-1} A_{31} \beta_{31}
\end{equation}
where $\beta_{31}$ is the escape probability for the photon emitted in
the decay 2p$^{3}$P$\rightarrow$ 1s$^{1}$S and $F_{\nu}^{C}$ is the
flux of the continuum at $\lambda$=10830~\AA, determined from the
observed infrared spectra. A($\lambda$) is the extinction calculated
using the empirical formula of Cardelli {\it et al.} (1989) assuming
A$_{V}$=0.6 (Blanco {\it et al.} 1987). For the distance D, a value of
50 kpc was adopted. 

For a deposition rate $\epsilon_{i}$ (erg cm$^{-3}$ s$^{-1}$) 
in species i, the abundance of the next ionisation stage i+1 is given by
the energy balance : \begin {equation}
n_{i+1}=\frac{\epsilon_{i}}{n_{e} \alpha_{i} w_{i}} \end {equation}
where $n_{e}$ is the electron density,
$\alpha_{i}$ is the recombination rate to all levels of the species i 
and $w_{i}$ is the energy required to produced an ion-electron pair.
Considering an envelope that consists only of hydrogen and helium and
taking the Penning ionisation into account, equation (14) reduces to: 
\begin{equation}
\frac{\epsilon_{H}}{w_{H}}+n_{He(2^{3}S)}n_{H~I}C_{P}=n_{H
II}n_{e}\alpha_{H~I}
\end{equation}
\begin{equation}
\frac{\epsilon_{He}}{w_{He}}=n_{He~II}n_{e}\alpha_{He}
\end{equation}
where \begin{math}\epsilon=\epsilon_{He}+\epsilon_{H}\end{math} is the
energy deposited in the envelope. The ratio of the energy deposited in
H and He is $\epsilon_{He}/\epsilon_{H}=mY$, where $Y=n_{He}/n_{H}$.
From the Bethe-Bloch formula for energy loss of fast electrons in H/He
material, we find $m$ to be 1.7. Assuming $n_{e}=n_{He II} +n_{ H
II}$ we can then calculate the population of the 2s$^{3}$S level,
$n_{2s^{3}S}$(R) as a function of radius, R, using
equations (15), (16) and (12). This requires the energy deposition rate
$\epsilon(R)$ which we obtain from the radiative transport calculation
described in 2.1, the total number density
$n_{tot}(R)=n_{He~I}(R)+n_{H~I}(R)+n_{e}(R)$, and the relative
abundance of hydrogen and helium, $Y=\frac{n_{He II}(R)+n_{He
I}(R)}{n_{H II}(R)+n_{H I}(R)}$. The last two quantities were obtained
from detailed explosion models.

The calculated density profile of the population of the 2s$^{3}$S
level, $n_{2s^{3}S}$(R), was then fed into our spectral synthesis code
(Fassia {\it et al.} 1998), thus producing a model He~I 10830~\AA\
P~Cygni line profile. This was then compared with the observed profile.

\section{Results}
We first applied the technique described above to the
radially-symmetric, unmixed model 10H (Woosley 1988).  This model
represents the explosion of a blue supergiant consisting of a
6~M$_{\odot}$ helium core, and a 10~M$_{\odot}$ hydrogen envelope. The
model incorporated an explosion energy of $1.4\times~10^{51}$~ergs and
a $^{56}$Ni mass of 0.07~M$_{\odot}$.  Among the {\it unmixed} explosion
models, 10H was one of the most successful at reproducing the observed
UVOIR light curve.  Nevertheless, significant discrepancies existed.
Model 10H also failed to explain the early detection of X-rays and
$\gamma$-rays.  Nor did it account for the high velocities observed in
the iron-group line profiles.  The present work shows (Figure~2a) that
model~10H also fails to reproduce the pronounced He~I absorption trough
observed at $\sim$10700~\AA. The lack of this He~I trough means that
the emission component of the C~I 10695~\AA\ line is unsuppressed
resulting in a poor match to the observed spectrum. This discrepancy is
not surprising since, in 10H, $^{56}$Ni only extends to 1100~km/s
(Figure~3), whereas at 76~days for example, the helium photosphere is
at $\sim4700$~km/s. The $\gamma$-rays simply do not penetrate to
sufficiently high velocities. 

In order to account for the early detection of the X-rays and
$\gamma$-rays Pinto \& Woosley (1988) introduced {\it ad hoc} outward
microscopic mixing into model 10H, yielding model 10HMM. In this model,
radioactive material is mixed outwards through the helium core and into
the hydrogen envelope. Model 10HMM successfully reproduced the UVOIR
light curve and accounted for the early and prolonged detection of
X-rays and $\gamma$-rays. However, on applying our technique to this
model, we found that it provides a poor match to the observed He~I
10830~\AA\ line profile (Figure~2b). This is because the extensive
dredge-up of $^{56}$Ni invoked in model 10HMM (Figure~3) produces a
2s$^{3}S$ population density that is much too high in the line forming
region.

In order to attempt to match the observed line profile we repeated the
microscopic mixing procedure that Pinto \& Woosley applied to model 10H
(Pinto, private communication).  Turbulent or macroscopic mixing will
be discussed in the next section. Starting from the centre, we
specified a velocity interval and homogenized the material composition
over that interval. We repeated this procedure, moving outwards until
the edge of the ejecta was reached.  This brings a fraction of the
radioactive material to high velocities. Nevertheless, the
abundance gradient is such that the bulk of the radioactive material
remains at low velocities.

For each of the mixed models produced we calculated the $\gamma$-ray
energy deposition, ionisation balance and 2s$^{3}S$ population density
as a function of radius, and then compared the resulting synthetic He~I
10830~\AA\ line profile with the observations.  Figure~3 shows the
$^{56}$Ni distribution for two of our mixed models (denoted 10HMA and
10HMB) that provided a good match to the observed He~I 10830 \AA\ line
profile. The synthetic helium line profiles deduced using model 10HMB
are compared with the observations in Figure~4. Equally good fits were
obtained for model 10HMA.

We also modelled the $\gamma$-ray line light curves for models 10HMA
and 10HMB using the radiative transport calculation explained in 2.1
and adopting $\kappa_{\nu}=N_{e}\times\sigma_{KN}(E_{0})$.  In
Figure~5, these light curves are compared with
$\gamma$-ray observations (Leising \& Share 1990; Sandie {\it et al.}
1988), assuming a distance of 50~kpc.  Also shown are the $\gamma$-ray
light curves deduced using model~10HMM.
 

\section{Discussion}
The results show that we can reproduce the He~I line by reducing the
amount of microscopic mixing that was invoked in model~10HMM. Indeed, in
model~10HMA there was negligible radioactive material at velocities
exceeding $\sim$3,900~km/s.  This is in contrast with 10HMM where
$\sim$1\% of the total $^{56}Ni$ mass has velocities in excess of
3,900~km/s. We find better agreement with the more recent work of
Nomoto {\it et al.} (1998). Based on their explosion model 14E13, their
bolometric light curve calculations suggest that $^{56}$Ni is mixed up to
3,000--4,000~km/s.

For models 10HMA and 10HMB we find that, respectively, 4\% and 3\% of
the total $^{56}$Ni mass (0.07 M$_{\odot}$) lay above 3,000~km/s. This
is consistent with late-time infrared spectroscopic observations which
showed that a small fraction of the cobalt and iron was travelling at
velocities $\sim$3,000~km/s (Spyromilio, Meikle \& Allen 1990; Haas {\it
et al.} 1990; Tueller {\it et al} 1990). From their 18~$\mu$m and
26~$\mu$m [Fe~II] line profiles, Haas {\it et al.} inferred that
$\geq4$\% of the iron mass had an expansion velocity greater than
3,000~km/s.  We also note that $\gamma$-ray light curve predictions
based on models 10HMA and 10HMB are consistent with the observations
(Figure~5).

Models 10HMM, 10HMA and 10HMB are equally successful at accounting for
the $\gamma$-ray light curves and late-time infrared line velocities.
However, in reproducing the observed early-time He~I 10830~\AA\ line,
models 10HMA and 10HMB are clearly superior to 10HMM. We conclude that,
given current technology, the He~I line technique provides a better
measure of the dredge-up of radioactive materials in type~II
supernovae.  Some of the advantage of the He~I line method stems from
the fact that it uses near-IR observations taken at relatively early
times when the supernova is bright.  An additional advantage is that,
since the method is based primarily on the optical depth of the He~I
P~Cygni line, it is insensitive to fluxing errors (Fassia {\it et al.}
1998).

In the above work, microscopic mixing was assumed, and so Penning
ionisation dominated the depopulation of the 2s$^{3}$S level 
({\it cf}~Fassia {\it et al.} 1998). However, hydrodynamic calculations
show that 
mixing is turbulent and takes place on macroscopic length scales.
Multi-dimensional studies predict that hydrogen and helium bubbles will
be dragged towards the inner parts of the ejecta, while clumps of
helium and heavier elements, including radioactive nickel, will
penetrate the hydrogen envelope (Fryxell, M\"{u}ller and Arnett 1991,
Herant \& Benz 1991).  Pure helium clumps in the hydrogen envelope will
tend to dominate the population of the 2s$^{3}$S level as they are not
subject to the effects of Penning ionisation.  Helium that is
microscopically mixed with hydrogen will make a negligible contribution
even if its abundance is much greater.  In Fassia {\it et al.} (1998)
we demonstrated that the presence of helium clumps in the hydrogen
envelope {\it reduces} significantly the degree of $^{56}$Ni dredge-up
that is needed to drive the He~I 10830~\AA\ line.  In that study we
assumed that the $^{56}$Ni was microscopically mixed with the helium
bubbles and the hydrogen envelope. However, in the case of SN~1987A,
there is good evidence that some of the $^{56}$Ni was ejected in
fast-moving clumps ({\it e.g.} Spyromilio, Meikle \& Allen 1990).  Thus,
if the $^{56}$Ni in the vicinity of the He~I line forming region is not
microscopically mixed, it may be necessary to {\it increase} the degree
of $^{56}$Ni dredge-up to provide the required rate of helium
ionisation.  Clearly, this is a complicated problem, requiring the use
of 2-D and 3-D explosion models. We therefore defer to a later paper an
examination of the implications of macroscopic mixing in SN~1987A for
the He~I technique. This should ultimately allow us to place
constraints on the initial inhomogeneities.

Finally, we note that the He~I observations take place well before the
$\gamma$-ray lines reach maximum brightness.  Thus, by observing the
He~I 10830~\AA\ line in type~II supernovae at early times, we should be
able to make accurate predictions of the $\gamma$-ray fluxes, providing
a valuable ``early-warning'' facility for $\gamma$-ray observers.\\

\section*{Acknowledgments}
We thank Phil Pinto and Stan Woosley for their advice and the use of
their explosion models. AF is supported by a scholarship from the
Alexander S Onassis Public Benefit Foundation.

\newpage 
\bf\Large{Figure Captions}\\
\\ 
\normalsize
Figure 1 : The coordinate system (p,z) used for the radiative transfer
calculation. The dashed line shows an example of an impact parameter
ray along which the frequency dependent equations of transfer were
integrated.

Figure 2 : Comparison of observed spectra of SN~1987A at 105~days, with
synthetic spectra whose He~I~10830 \AA\ line profiles are based on
models 10H and 10HMM. Data are from Elias {\it et al.} (1988).

Figure 3 : Distributions of $^{56}$Ni for models 10H (dashed line), 10HMM
(dotted line), 10HMA (solid line) and 10HMB (dot-dashed line). 

Figure 4 : Comparison of observed spectra with synthetic spectra whose
He~I~10830 \AA\ line profiles are based on model 10HMB.  Data are from
Elias {\it et al.} (1988) and Meikle {\it et al.} (1989).

Figure 5 : Comparison of the $\gamma$-ray line light curves of
model~10HMA (solid line) and model~10HMB (dotted line) with the
observations. Data are from Leising 
\& Share (1990) (open triangles) and from Sandie {\it et al.}(1988)
(open squares).  The dashed line represents the predictions of 
model~10HMM.
\newpage
\begin{figure}
\epsfysize=0.65\textwidth \epsfbox{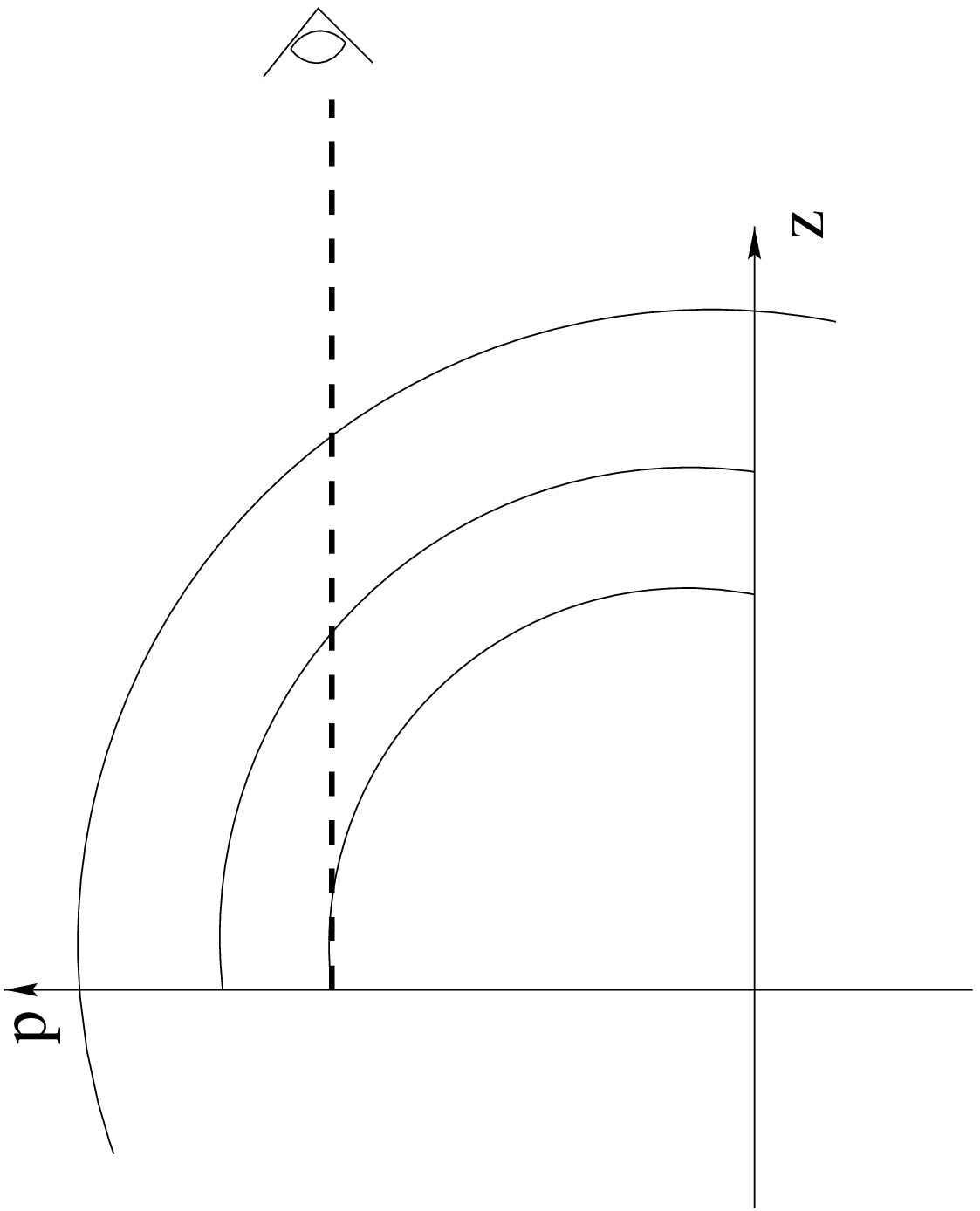}
\caption{}
\end{figure}
\begin{figure} 
\epsfxsize=0.85\textwidth \epsfbox{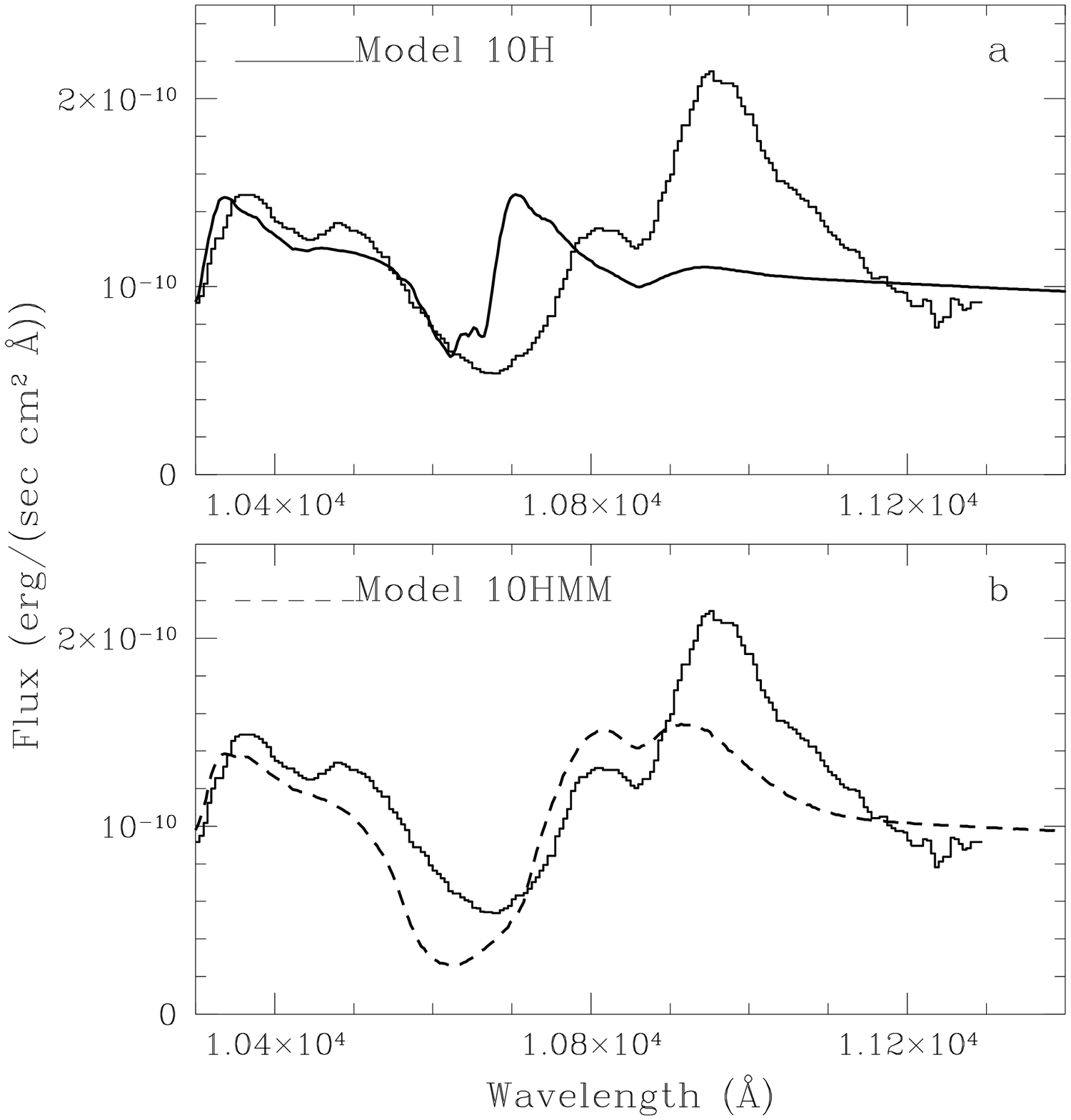}
\caption{}
\end{figure}
\begin{figure}
\epsfxsize=0.85\textwidth \epsfbox{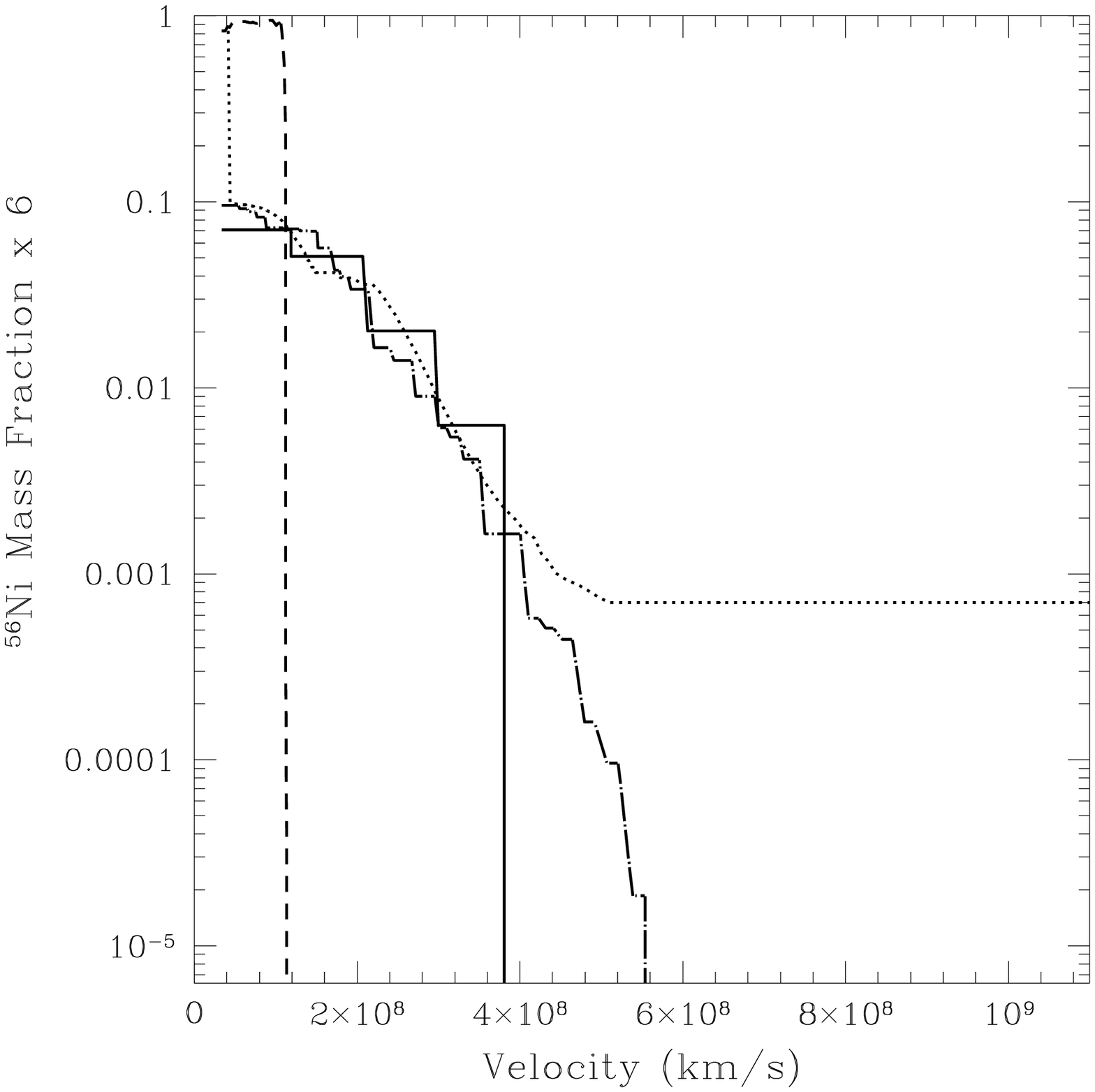}
\caption{}
\end{figure}
\begin{figure}
\epsfxsize=0.85\textwidth \epsfbox{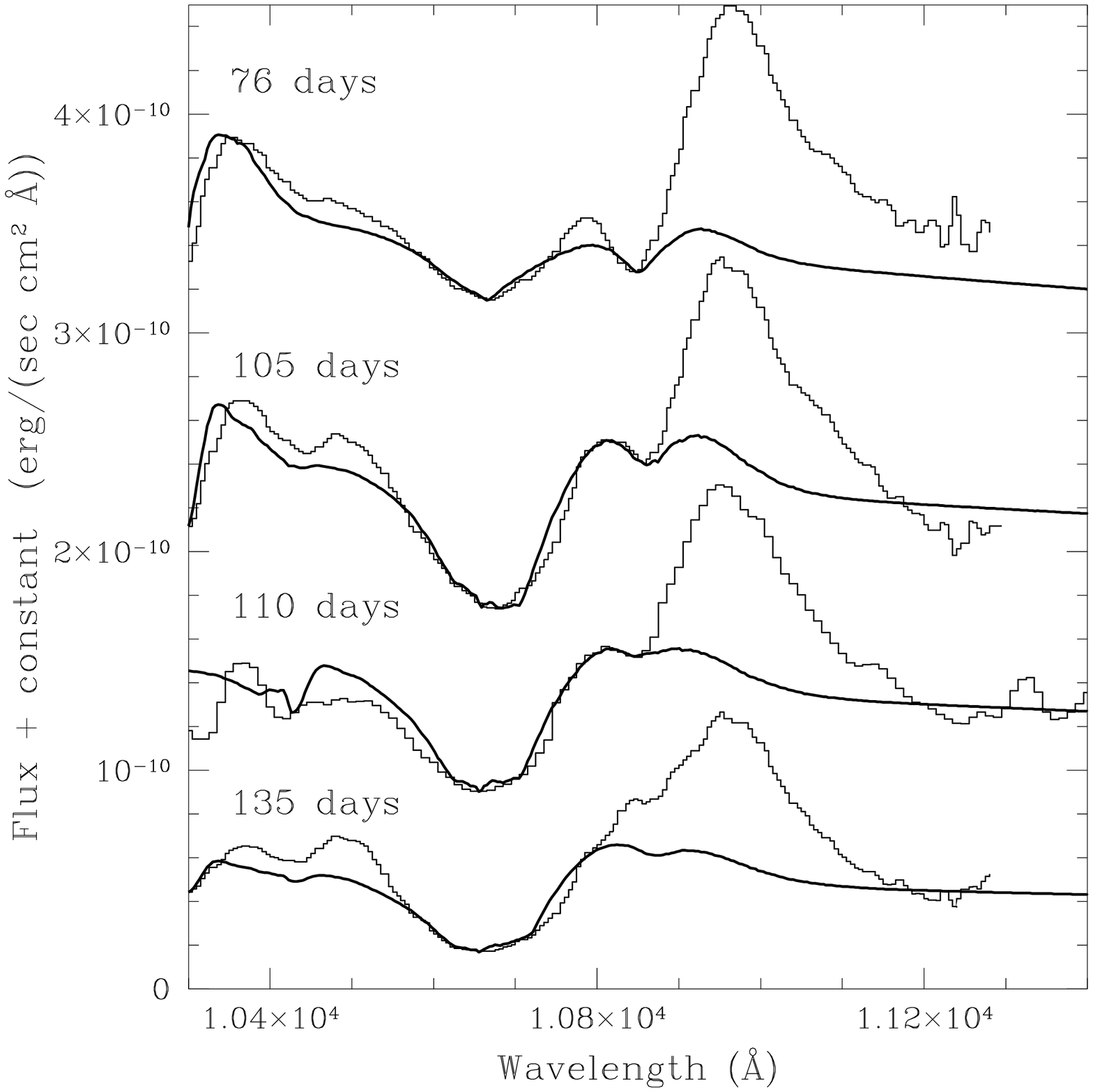}
\caption{}
\end{figure}
\begin{figure}
\epsfxsize=0.85\textwidth \epsfysize=3.9in \epsfbox{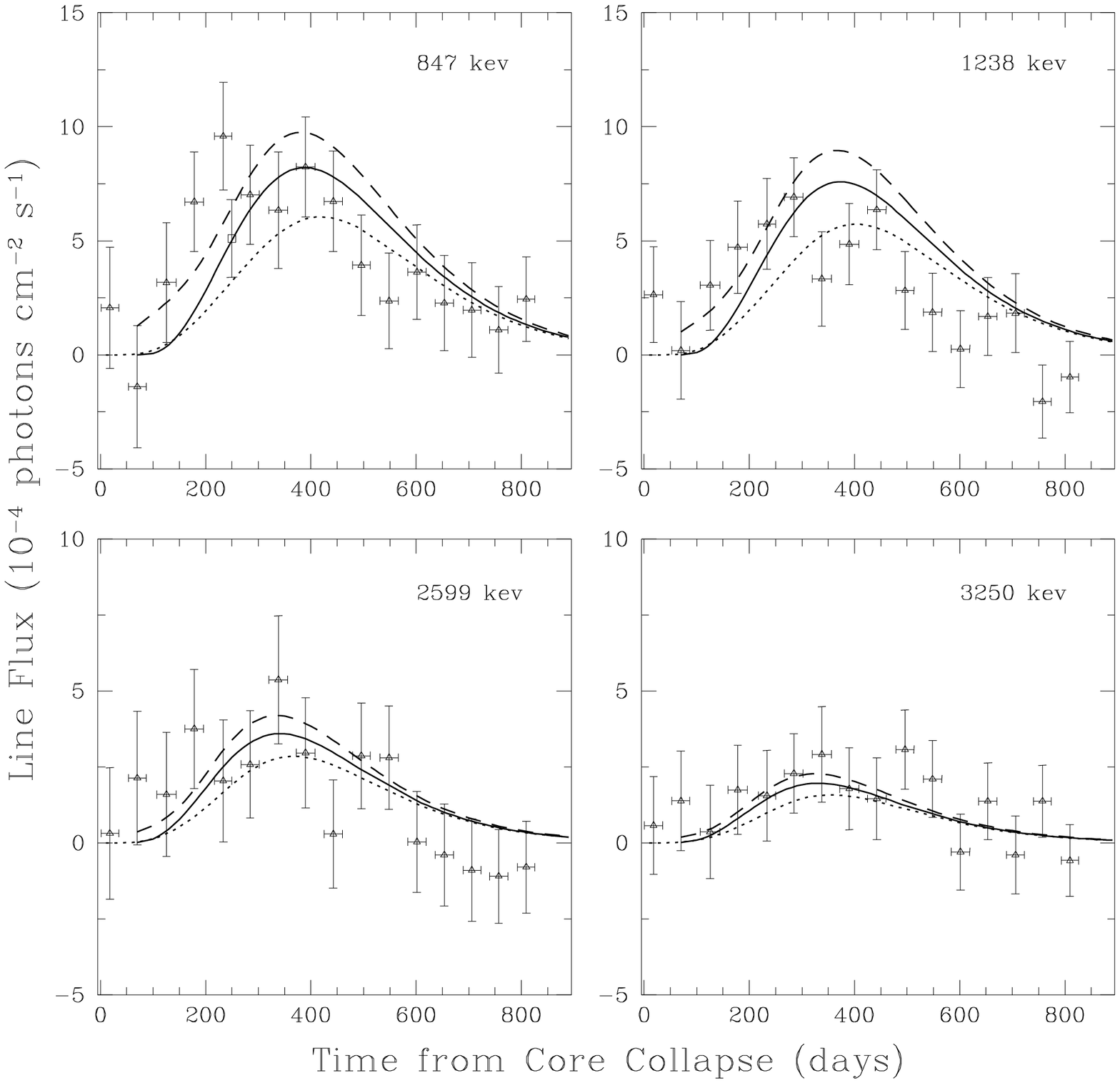}
\caption{}
\end{figure}
\end{document}